\documentstyle[aps,prl,epsf,multicol,amsmath,graphics,psfrag,%
  graphicx,pstcol,psfig]{revtex}

\begin{document}
\draft

\title{
  Charge excitations in LiV$_{2}$O$_{5}$ and NaV$_{2}$O$_{5}$: similarities 
  and differences
}

\author{A. H\"{u}bsch, C. Waidacher, and K. W. Becker}
\address{
  Institut f\"{u}r Theoretische Physik,
  Technische Universit\"{a}t Dresden, D-01062 Dresden, Germany
}

\date{\today}
\maketitle

\begin{abstract}
  We calculate the optical conductivity of LiV$_{2}$O$_{5}$ and 
  NaV$_{2}$O$_{5}$ using exact numerical diagonalization of a quarter-filled 
  extended Hubbard model on a system of coupled ladders. In particular, 
  electronic correlations are treated exactly, and a quantitative agreement 
  between calculated and experimentally observed optical conductivity of these 
  two vanadium oxides is presented. Furthermore, it is found that 
  LiV$_{2}$O$_{5}$ differs from NaV$_{2}$O$_{5}$ not only in the charge 
  ordering pattern but also in the nature of the inter-ladder coupling: In 
  contrast to LiV$_{2}$O$_{5}$, in NaV$_{2}$O$_{5}$ neighboring ladders are 
  coupled by a strong Coulomb repulsion, and not by inter-ladder hopping.
\end{abstract}

\widetext

\begin{multicols}{2}
\narrowtext


In the past years, low-dimensional transition metal compounds have been 
intensively investigated because of their unconventional spin and charge 
excitation spectra \cite{Dagotto}. In this respect, the vanadate 
$\alpha'$-NaV$_{2}$O$_{5}$ has attracted particular interest as a ladder 
system at quarter filling, containing only one equivalent V site with a formal 
valence $+4.5$ \cite{Smolinski,Horsch}. The magnetic susceptibility of 
NaV$_{2}$O$_{5}$ can be well described by a $S=1/2$ antiferromagnetic 
Heisenberg chain with exchange interactions of $J=440$ and $560$~K for 
temperatures below \cite{Weiden} and above \cite{Isobe} the transition 
temperature $T_{C}\approx 34$~K. The low-temperature phase is found to be 
charge ordered \cite{Ohama}, but the nature of this transition is still under 
discussion \cite{Gros,Trebst,Bernert}. The much less studied 
$\gamma$-LiV$_{2}$O$_{5}$ belongs to the same family of vanadium oxides and 
exhibits a one-dimensional $S=1/2$ Heisenberg like behavior with an exchange 
interaction of $J=308$~K \cite{Isobe_96,Fujiwara_97}. In contrast to 
NaV$_{2}$O$_{5}$, there is no indication of a phase transition at lower 
temperature. Since both compounds are structurally related it is interesting 
to clarify the microscopic origin of the different physical properties of both 
vanadates. This will be discussed in this paper on the basis of the optical 
conductivity.

Recently, the optical conductivity of LiV$_{2}$O$_{5}$ and NaV$_{2}$O$_{5}$ 
has been measured \cite{Konstantinovic}. In the energy range from $0$ to 
$3$~eV similar peaks in the ${\bf E}\parallel a$ spectra of both materials 
were found. On the other hand, a complete suppression of the peaks in the 
${\bf E}\parallel b$ spectrum of LiV$_{2}$O$_{5}$ was 
observed. In this paper, we show that this suppression results not only from 
the double chain charge ordering pattern but also from the strong inter-ladder 
hopping in LiV$_{2}$O$_{5}$. Consequently, it is found that both vanadates 
differ not only in the charge ordering pattern but also in the nature of the 
inter-ladder coupling. Recently, we studied \cite{Huebsch} the optical 
conductivity of NaV$_{2}$O$_{5}$ so that in the present paper we concentrate 
on LiV$_{2}$O$_{5}$ and on the comparison of both vanadium oxides.

\begin{figure}
  \begin{center}
    \scalebox{0.5}{
      \includegraphics*[100,310][480,740]{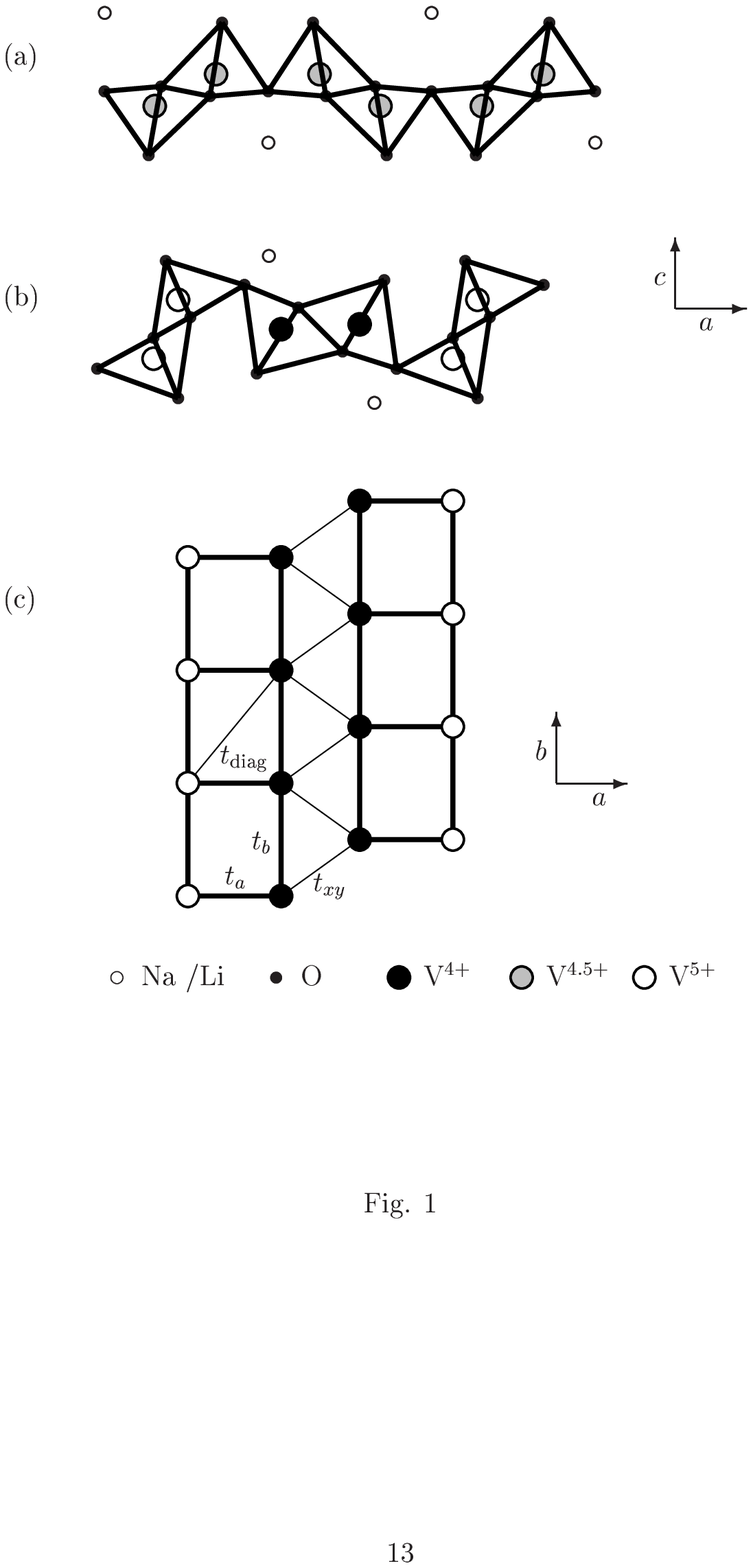}
    }
  \end{center}
  \caption{
    Representation of the (a) NaV$_{2}$O$_{5}$ and (b) LiV$_{2}$O$_{5}$ room 
    temperature crystal structure. (c) shows a schematic projection of the 
    LiV$_{2}$O$_{5}$ structure onto a ($a$,$b$) plane where only the vanadium 
    ions are drawn.
  }\label{structure}
\end{figure}

At room temperature LiV$_{2}$O$_{5}$ and NaV$_{2}$O$_{5}$ have orthorhombic 
crystal structures that are described by space groups $Pnma$ and $Pmmn$, 
respectively \cite{Anderson,Schnering}. Both compounds consist of layers of 
VO$_{5}$ square pyramids (see Fig.~\ref{structure}). In contrast to 
NaV$_{2}$O$_{5}$, a structural analysis \cite{Anderson} for LiV$_{2}$O$_{5}$ 
shows two inequivalent vanadium sites which were also found by NMR 
experiments \cite{Fujiwara_98}. These two vanadium sites were assigned a 
valence of V$^{4+}$ and V$^{5+}$, respectively, and they form two different 
zig-zag chains along the $b$ direction \cite{Anderson}. Therefore, magnetic 
V$^{4+}$ double chains are separated by nonmagnetic V$^{5+}$ double chains. 
Note that this charge ordering pattern is not comparable with the in-line 
order discussed for NaV$_{2}$O$_{5}$ \cite{Thalmeier}. As can be seen from 
Fig.~\ref{structure}, the VO layers are more corrugated in LiV$_{2}$O$_{5}$ 
than in NaV$_{2}$O$_{5}$ because the Li atoms are smaller than the Na atoms.


In Ref.~\onlinecite{Huebsch} we studied the optical conductivity of 
NaV$_{2}$O$_{5}$ using a quarter-filled extended Hubbard model of the V $3d$ 
electrons on the vanadium lattice which forms a coupled ladder system. In the 
present study we modify this model for LiV$_{2}$O$_{5}$ by an additional 
on-site energy $\varepsilon_{i}$
\begin{eqnarray}
  {\cal H} &=& -\sum_{\langle i,j\rangle,\sigma}t_{ij}
               \left(
                 c_{i,\sigma}^{\dagger}c_{j,\sigma}+{\rm H.c.}
               \right)
               + U\sum_{i}n_{i\uparrow}n_{i\downarrow}\nonumber\\
  &&           +\sum_{\langle i,j\rangle}V_{i,j}n_{i}n_{j} 
               +\sum_{i}\varepsilon_{i}\,n_{i}\label{hamilton}
\end{eqnarray}
to cause the double chain charge ordering observed in 
LiV$_{2}$O$_{5}$ \cite{Anderson}. In Eq.~\eqref{hamilton} $\langle i,j\rangle$ 
denotes summation over all pairs of nearest neighbors, and spin 
$\sigma=\uparrow,\downarrow$. $c_{i,\sigma}^{\dagger}$ are electron
creation operators at vanadium sites, 
$n_{i}=\sum_{\sigma}c_{i,\sigma}^{\dagger}c_{i,\sigma}$ is
the occupation-number operator, and $U$ denotes the Coulomb repulsion between
electrons on the same site. The hopping parameters $t_{ij}$ of 
Eq.~\ref{hamilton} are defined in panel (c) of Fig.~\ref{structure}. The 
inter-site Coulomb interactions $V_{ij}$ between nearest vanadium neighbors on 
a rung, on a leg or on different ladders are $V_{a}$, $V_{b}$ and $V_{xy}$. 
The on-site energy $\varepsilon_{i}$ is $-\Delta/2$ ($\Delta/2$) for the 
V$^{4+}$ (V$^{5+}$ sites [see panel (c) in Fig.~\ref{structure}]. In contrast 
to NaV$_{2}$O$_{5}$, in LiV$_{2}$O$_{5}$ the ladders are less strictly 
situated in an $(a,b)$ plane (see Fig.~\ref{structure}). Thus, the system 
shown in panel (c) of Fig.~\ref{structure} has more 3D character than 
NaV$_{2}$O$_{5}$.

For an insulating system, the optical conductivity $\sigma_{\alpha}(\omega)$ 
is given by the Kubo formula for finite frequency response
\begin{eqnarray}
  \sigma_{\alpha}(\omega) &=& \frac{1}{\omega}{\rm Re}\int_{0}^{\infty}dt\,
    e^{i\omega t}\langle 0| j_{\alpha}(t)j_{\alpha}|0\rangle.\label{optic}
\end{eqnarray}
Here $j_{\alpha}$ with $\alpha=a,b$ are the components of the current
operator parallel to $a$ or $b$ direction. Equation \eqref{optic} is valid 
for zero temperature, whereas the experiments have been carried out at finite 
temperatures \cite{Konstantinovic}. However, it is known from both experiment 
and theory that at least the spectra for NaV$_{2}$O$_{5}$ depend only weakly 
on temperature \cite{Long,Presura}. Therefore, we may restrict ourselves to 
calculations for zero temperature.

In the following we evaluate Eq.~\eqref{optic} using the standard Lanczos 
algorithm\cite{Lin} which is limited to small clusters. Therefore, first we 
have to check if our results are sufficiently converged with respect to system 
size. Since the hopping between V$^{5+}$ ions in LiV$_{2}$O$_{5}$ is very 
small \cite{Valenti_01}, this material can be described by the double ladder 
system shown in panel (c) of Fig.~\ref{structure}. Consequently, the system is 
infinite only in $b$-direction, and we shall always use open boundary 
conditions for this direction (to enlarge the effective cluster size). 
However, one has to make sure that electrons on the edges of the cluster are 
still embedded in the local Coulomb potential that results from the double 
chain charge ordering. For this purpose, sites on the edges of the clusters 
that are occupied in the perfectly ordered state need an additional on-site 
energy $V_{b}$ to simulate the influence of occupied sites outside of the 
cluster. To prove that the results are sufficiently converged with respect to 
system size we have evaluated the double ladder system shown in panel (c) of 
Fig.~\ref{structure} both with open and periodic boundary conditions and found 
only small differences (not shown).

The distances within the vanadium ladder structures in LiV$_{2}$O$_{5}$ and 
NaV$_{2}$O$_{5}$ are similar. Therefore, we start from the parameters of the 
ladder used in Ref.~\onlinecite{Huebsch} for NaV$_{2}$O$_{5}$ 
($t_{a}=0.38$~eV, $t_{b}=0.17$~eV, $t_{\rm diag}=0$, $V_{a}=0.8$~eV, 
$V_{b}=0.6$~eV, $U=2.8$~eV) to describe the optical conductivity of 
LiV$_{2}$O$_{5}$. Consequently, we have to determine only three parameter 
values, $\Delta$, $V_{xy}$, and $t_{xy}$. For LiV$_{2}$O$_{5}$ the choice 
$V_{xy}=0$ is possible because the Coulomb repulsion between nearest neighbors 
of neighboring ladders counteracts the double chain charge ordering caused by 
$\Delta$. This means that the parameters $\Delta$ and $V_{xy}$ are not 
independent from each other.

\begin{figure}
  \begin{center}
    \scalebox{0.5}{
      \includegraphics*[110,370][480,770]{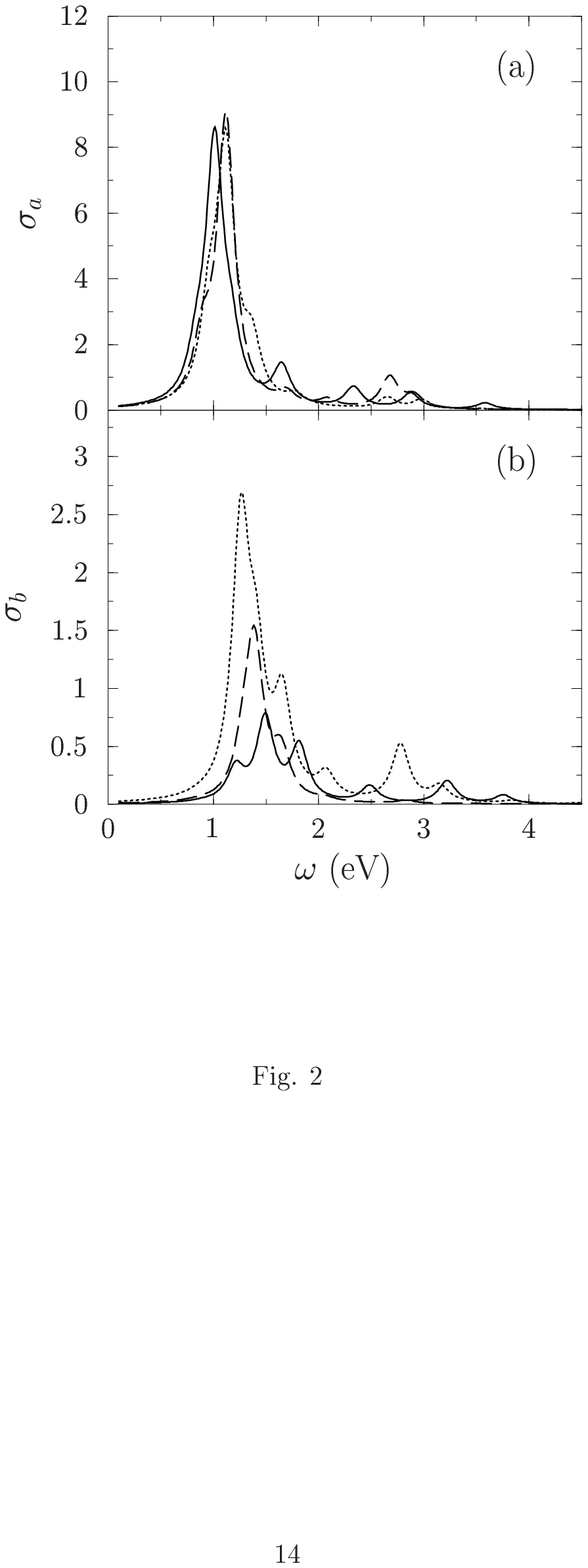}
    }
  \end{center}
  \caption{
    Optical conductivity [panel (a): $\sigma_{a}(\omega)$, panel (b): 
    $\sigma_{b}(\omega)$] of a single ladder ($t_{xy}=0$) where the hopping 
    parameters from Ref. [\ref{Smolinski}] (dotted lines, 
    $t_{a}=0.38$~eV, $t_{b}=0.17$~eV, $t_{\rm diag}=0$) and from Ref. 
    [\ref{Valenti_01}] (dashed lines, $t_{a}=0.35$~eV, $t_{b}=0.02$~eV, 
    $t_{\rm diag}=0.1$~eV) were used. In addition, the results for the double 
    ladder system shown in panel (c) of Fig.~\ref{structure} ($t_{a}=0.35$~eV, 
    $t_{b}=0.02$~eV, $t_{\rm diag}=0.1$~eV, $t_{xy}=-0.18$~eV) is plotted with 
    full lines. The parameters of the Coulomb interaction are $U=2.8$~eV, 
    $V_{a}=0.8$~eV, $V_{b}=0.6$~eV, $V_{xy}=0$, and $\Delta=1.7$~eV. The 
    theoretical line spectra are broadened with Gaussian function of width 
    $0.1$ eV.
  }\label{t_xy}
\end{figure}

First we discuss the case of a single ladder ($t_{xy}=0$) for which the 
optical conductivity is shown as dotted lines in Fig.~\ref{t_xy}. As can be 
seen from panel (a) of Fig.~\ref{t_xy}, $\sigma_{a}$ is dominated by an 
excitation at $1.0$~eV which corresponds to a transition from a bonding to an 
anti-bonding state of a singly occupied rung that has been discussed before 
for NaV$_{2}$O$_{5}$ \cite{Damascelli2000}. If one considers a single rung 
one obtains $\sqrt{\delta^{2}+4t_{a}^{2}}$ for the excitation energy of this 
transition where an effective on-site energy $\delta=\Delta-2V_{b}$ has to be 
used to take the chain charge ordering into account. Therefore, the on-site 
energy $\Delta$ determines the excitation energy (of the transition from a 
bonding to an anti-bonding state). A value of $\Delta=1.7$~eV can be obtained 
directly from the experimentally observed peak position \cite{Konstantinovic}. 
In contrast to $\sigma_{a}$, for $\sigma_{b}$ we find two basic mechanisms for 
excitations. The structures at $1.2$~eV and $1.8$~eV [see dotted line in panel 
(b) of Fig.~\ref{t_xy}] can be interpreted as a transition to states with one 
unoccupied and one doubly occupied rung [excitation energies 
$\sim(\Delta+V_{a}-2V_{b})$ and $\sim(\Delta+V_{a}-V_{b})$]. The excitations 
around $2.8$~eV result from the creation of one doubly occupied site 
(excitation energy $\sim U$).

Next we discuss the influence of the inter-ladder hopping $t_{xy}$ on the 
optical conductivity. A moderate value of $t_{xy}$ affects $\sigma_{a}$ 
only weakly and leads, on the other hand, to a suppression of low-energy peaks 
in $\sigma_{b}$ (not shown). This suppression becomes stronger if $\Delta$ 
increases. (Because of different charge ordering pattern this influence of 
$\Delta$ is not identical with the findings in Ref.~\cite{Nishimoto}.) One 
would obtain good agreement between the calculated and the experimentally 
observed \cite{Konstantinovic} optical conductivity for 
$t_{xy}=0.3\dots 0.4$~eV. However, according to Ref.~\onlinecite{Valenti_01}, 
the magnetic properties of the model along $b$-direction can be described by a 
single Heisenberg chain along a zig-zag line. The choise of 
$t_{xy}=0.3\dots 0.4$~eV for the inter-ladder hopping supposed above would 
lead to a strong frustration 
\cite{frust}
\begin{eqnarray}
  \frac{J_{b}}{J_{xy}} &=& \frac{t_{b}^{2}}{(U-V_{b})}\,
                           \frac{U}{t_{xy}^{2}}
                           \, \approx\, 0.3\label{frust}
\end{eqnarray}
which is not in accordance with NMR\cite{Fujiwara_97} and susceptibility 
measurements \cite{Isobe_96}. However, this discrepancy is caused by 
neglecting the diagonal hopping $t_{\rm diag}$: Tight-binding fits to the band 
structure found in LDA calculations with nonzero $t_{\rm diag}$ lead to a 
small value of $t_{b}$ both for NaV$_{2}$O$_{5}$ and for LiV$_{2}$O$_{5}$ 
\cite{Yaresko,Valenti_01}. Therefore, we use in the following the set of 
tight-binding parameters from Ref.~\onlinecite{Valenti_01} for 
LiV$_{2}$O$_{5}$ ($t_{a}=0.35$~eV, $t_{b}=0.02$~eV, $t_{\rm diag}=0.1$~eV, 
$t_{xy}=-0.18$~eV). Since the diagonal hopping $t_{\rm diag}$ is included, the 
frustration parameter \eqref{frust} is now negligibly small \cite{Valenti_01}. 
The Coulomb interactions $U$, $V_{a}$, and $V_{b}$ of the ladder are not 
changed. In contrast to LiV$_{2}$O$_{5}$, for NaV$_{2}$O$_{5}$ we use a 
parameter set with $t_{\rm diag}=0$ as in Ref.~\onlinecite{Huebsch} to avoid 
strong finite-size effects \cite{Note}.

From a comparison of the optical conductivity of a single ladder for both 
parameter sets (dotted and dashed lines in Fig.~\ref{t_xy} one can conlude 
that the results are not fundamentally changed by the tight-bindung parameters 
from Ref.~\onlinecite{Valenti_01}. But these new parameters lead to a 
significantly decreased spectral weight of the low-energy structure in 
$\sigma_{b}$ because the mobility of a created doubly occupied rung is now 
more restricted due to the small value of $t_{b}$. Nevertheless, the above 
analysis of the basic mechanisms for excitations on a single ladder remains 
valid even for $t_{\rm diag}\not= 0$. 

Now we return to the influence of the inter-ladder hopping on the optical 
conductivity. For this purpose, we compare in Fig.~\ref{t_xy} the results of a 
single ladder (dashed lines) with the optical conductivity for the complete 
double ladder system shown in panel (c) of Fig.~\ref{structure} (full lines) 
where we use the hopping parameters from Ref.~\onlinecite{Valenti_01}. As can 
be seen from panel (a) the excitation resulting from the transition between a 
bonding and an anti-bonding state of a rung discussed above depends only 
weakly on $t_{xy}$. In contrast to $\sigma_{a}$, $\sigma_{b}$ is changed 
strongly by a finite $t_{xy}$. In particular, some new structures arise, and 
the spectral weight is divided among different excitations. Note that the 
influence of $t_{xy}$ does not depend on its sign.

\begin{figure}
  \begin{center}
    \scalebox{0.5}{
      \includegraphics*[100,430][520,795]{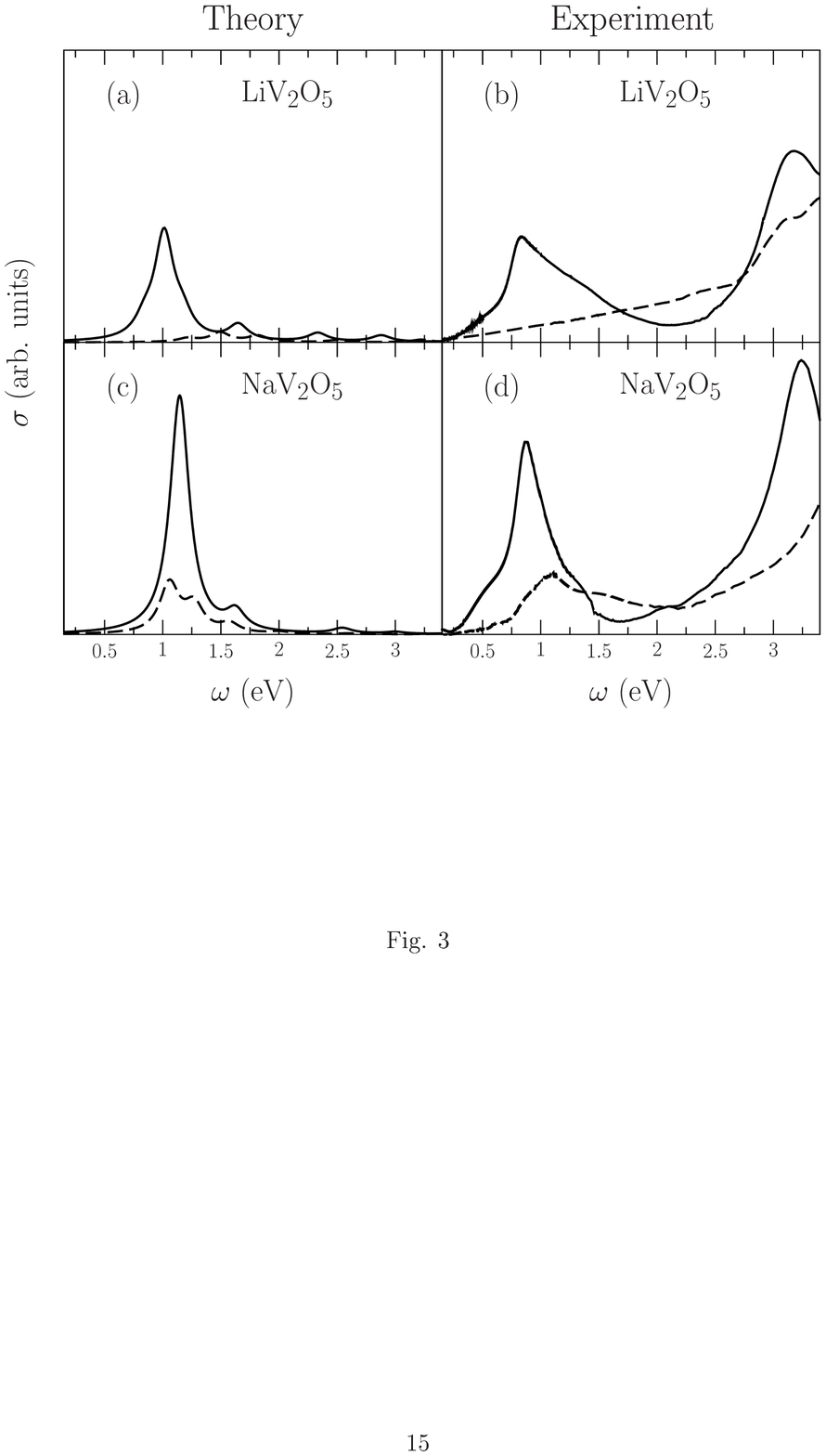}
    }
  \end{center}
  \caption{
    Comparison of experimental data for LiV$_{2}$O$_{5}$ [panel (b)] and for 
    NaV$_{2}$O$_{5}$ [panel (d)], taken from Ref.~\ref{Konstantinovic}, 
    and the results of the exact diagonalization [panels (a) and (c)] whereby  
    $\sigma_{a}$ ($\sigma_{b}$) is plotted with full (dashed) lines. The 
    hopping parameters are taken from Ref.~\ref{Valenti_01} (LiV$_{2}$O$_{5}$) 
    and Ref.~\ref{Smolinski}, respectively. The Coulomb interactions on the 
    ladders ($U=2.8$~eV, $V_{a}=0.8$~eV, $V_{b}=0.6$~eV) are the same for both 
    compounds. Additional parameters are $\Delta=1.7$~eV for LiV$_{2}$O$_{5}$ 
    and $V_{xy}=0.9$~eV for NaV$_{2}$O$_{5}$. The theoretical line spectra 
    have been convoluted with a Gaussian function of width $0.1$~eV.
  }\label{comp}
\end{figure}

In Fig.~\ref{comp} the calculated optical conductivities for LiV$_{2}$O$_{5}$ 
and NaV$_{2}$O$_{5}$ [panels (a) and (c)] are compared to the experimental 
spectra [panels (b) and (d)] from Ref.~\onlinecite{Konstantinovic}. We 
restrict the following discussion of the optical conductivity to the energy 
range up to $2$~eV since transitions with a higher excitation energy involve 
oxygen orbitals \cite{Konstantinovic}. Good agreement for the optical 
conductivity of NaV$_{2}$O$_{5}$ is found [compare panels (c) and (d) of 
Fig.~\ref{comp}] where the hopping parameters ($t_{a}=0.38$~eV, 
$t_{b}=0.17$~eV, $t_{xy}=0.012$eV) and the on-site Hubbard interaction 
($U=2.8$~eV) are taken from Ref.~\onlinecite{Smolinski}. The values of the 
intersite Coulomb interactions $V_{a}=0.8$~eV, $V_{b}=0.6$~eV and 
$V_{xy}=0.9$~eV have been adjusted to obtain correct peak positions (for 
details see Ref.~\onlinecite{Huebsch}). Note that the loss function in EELS 
experiments for NaV$_{2}$O$_{5}$ can also be described with the same 
Hamiltonian, and the same set of model parameters \cite{Huebsch}. For 
LiV$_{2}$O$_{5}$ we obtain good agreement of the calculated optical 
conductivity [panel (a) of Fig.~\ref{comp}] and the experimental data of 
Ref.~\onlinecite{Konstantinovic} [panel (b) of Fig.~\ref{comp}]. Again we use 
the hopping parameters from LDA calculations\cite{Valenti_01} 
($t_{a}=0.35$~eV, $t_{b}=0.02$~eV, $t_{\rm diag}=0.1$~eV, $t_{xy}=-0.18$~eV) 
and the Coulomb interactions of the ladder ($U=2.8$~eV, $V_{a}=0.8$~eV, 
$V_{b}=0.6$~eV) found for NaV$_{2}$O$_{5}$. The on-site energy $\Delta=1.7$~eV 
is adjusted to obtain the correct peak position of $\sigma_{a}$. Thus we use 
only one free parameter for LiV$_{2}$O$_{5}$. The effective on-site energy 
$\delta=\Delta-2V_{b}=0.5$~eV that determines the charge ordering along the 
chains agrees well with the value $\delta=0.3$~eV found in 
Ref.~\onlinecite{Valenti_01}. To summarize the comparison of the calculated 
and the experimentally observed optical conductivity for LiV$_{2}$O$_{5}$ and 
NaV$_{2}$O$_{5}$ (Fig.~\ref{comp}) one can state that a good agreement is 
found for both vanadates. In particular, the complete suppression of the low 
energy structure in $\sigma_{b}$ of LiV$_{2}$O$_{5}$, and the peak shapes are 
reproduced.

Finally, we want to compare the results for LiV$_{2}$O$_{5}$ and 
NaV$_{2}$O$_{5}$ in some detail. As shown above, the ladder parameters are 
similar for both compounds. This means that the electronic properties of the 
single ladders are similar in LiV$_{2}$O$_{5}$ and NaV$_{2}$O$_{5}$ even 
though one finds a double chain charge ordering in LiV$_{2}$O$_{5}$ caused by 
an on-site energy $\Delta$. Apart from that, the main difference between 
LiV$_{2}$O$_{5}$ and NaV$_{2}$O$_{5}$ is the character of the inter-ladder 
coupling. In LiV$_{2}$O$_{5}$ one finds a strong inter-ladder hopping 
$t_{xy}$ [keep in mind that this coupling does not go beyond the double ladder 
systems shown in panel (c) of Fig.~\ref{structure}]- which is very small in 
NaV$_{2}$O$_{5}$. On the other hand, in NaV$_{2}$O$_{5}$ the ladders are 
coupled due to a strong inter-ladder Coulomb interaction $V_{xy}$. Note again 
that the choise $V_{xy}=0$ is reasonable for LiV$_{2}$O$_{5}$ since $\Delta$ 
and $V_{xy}$ are not independent from each other. The character of the 
inter-ladder coupling determines the low energy features of $\sigma_{b}$: the 
well defined structure in NaV$_{2}$O$_{5}$ is suppressed in LiV$_{2}$O$_{5}$ 
(see Fig.~\ref{comp}) where a finite $t_{xy}$ leads to new excitations among 
which the spectral weight is divided.

In conclusion, we have obtained a quantitative description of the highly 
anisotropic optical conductivity of LiV$_{2}$O$_{5}$ and NaV$_{2}$O$_{5}$ by 
exact diagonalization of small clusters. In particular, the complete 
suppression of the low-energy peaks in the spectrum of LiV$_{2}$O$_{5}$ with 
electrical field parallel to the leg direction is reproduced. The good 
agreement of calculated and experimental optical conductivity for both 
vanadates implies that the electronic properties of the single ladders are 
similar in LiV$_{2}$O$_{5}$ and NaV$_{2}$O$_{5}$ (even though one finds a 
double chain charge ordering in LiV$_{2}$O$_{5}$). The main differences 
between the optical properties of LiV$_{2}$O$_{5}$ and NaV$_{2}$O$_{5}$ are 
due to the completely different nature of the inter-ladder coupling. In 
contrast to LiV$_{2}$O$_{5}$, in NaV$_{2}$O$_{5}$ different ladders are 
coupled by a strong Coulomb repulsion $V_{xy}$ and not by inter-ladder hopping 
$t_{xy}$. Because of the on-site energy $\Delta$ LiV$_{2}$O$_{5}$ can be 
viewed as an asymmetric quarter-filled ladder compound. Therefore, it would be 
interesting to compare with other experiments probing charge excitations like 
EELS or XPS.


We would like to acknowledge fruitful discussions with S. Atzkern, J. Fink, 
M. S. Golden, R. E. Hetzel, and W. von der Linden. Furthermore, we thank 
M. J. Konstantinovi\'{c} for sending us his experimental data. This work was 
supported by DFG through the research program GK 85, Dresden.



\end{multicols}
\widetext


\begin{references}
\bibitem{frust}$J_{xy}$ is the Heisenberg exchange between adjacent spins of 
    different ladders, $J_{b}$ couples spins on a rung. For a single 
    Heisenberg chain along a zig-zag line $J_{xy}$ ($J_{b}$) has to be 
    interpreted as the exchange bewteen (next) nearest neighbor spins.
\bibitem{Note} To calculate the optical conductivity for NaV$_{2}$O$_{5}$ 
    reliably we have studied two different adapted clusters for the electric 
    field parallel or perpendicular to the leg direction \cite{Huebsch}. Note, 
    for finite diagonal hopping $t_{\rm diag}$ these clusters can not be used.
\end{references}
\end{document}